%% file: ctrundle08_1.tex
\documentclass[oldversion]{aa}
\usepackage{graphicx}
\usepackage{times}
\usepackage{times,graphics}
\usepackage{lscape}
\usepackage{longtable}
\usepackage{epsfig} 
\usepackage{txfonts} 
\usepackage{natbib}
\usepackage{psfig}
\bibpunct{(}{)}{;}{a}{}{,}

\def\5{\footnotesize V\normalsize}
\def\4{\footnotesize IV\normalsize}
\def\3{\footnotesize III\normalsize}
\def\2{\footnotesize II\normalsize}
\def\1{\footnotesize I\normalsize}

\def\kms{$\mbox{km s}^{-1}$}

\begin{document}

\title{SN 2005 gj: Evidence for LBV supernovae progenitors? \thanks{Based on observations at the European Southern Observatory with UVES
on the VLT in programme 276.D-5020A and 278.D-0270.}}
   \author{C. Trundle \inst{1}, R. Kotak \inst{1}, J.S. Vink\inst{2}, W.P.S. Meikle\inst{3}} 

   \offprints{C.Trundle,~\email{c.trundle@qub.ac.uk.}}

   \authorrunning{C.Trundle}
   
   \titlerunning{SN 2005gj: Evidence for LBV supernovae progenitors? }
    
   \institute{Astronomy Research Centre, Department of Physics \& Astronomy,
   School of Mathematics \& Physics, The Queen's University of Belfast, Belfast,
   BT7 1NN, Northern Ireland 
    \and                
       Armagh Observatory, College Hill, Armagh, BT61 9DG, Northern Ireland
   \and                
      Astrophysics group, Imperial College London, Blackett Laboratory, Prince Consort Rd, London, SW7 2BZ}
   \date{}

   \abstract{There has been mounting observational evidence in favour of Luminous 
Blue Variables (LBVs) being the direct progenitors of supernovae. Here we present
possibly the most convincing evidence yet for such progenitors. We find multiple
absorption component P-Cygni profiles of hydrogen and helium  in the spectrum of
SN~2005gj, which we interpret as being an imprint of  the progenitors mass-loss
history.   Such profiles have previously only been detected in Luminous Blue
Variables. This striking resemblance of the profiles, along with wind velocities and
periods consistent with LBV's leads us to connect SN~2005gj to an LBV progenitor.} %
 \maketitle  


\section{Introduction} 

In the canonical picture of the evolution of very massive
($M\gtrsim40\,M_\odot$) stars \citep[e.g.][]{lan94}, the role of mass loss
is widely accepted as being the key process which drives O and B-type stars 
into the luminous blue variable (LBV) phase, before entering the He-burning
Wolf-Rayet (WR) stage, at the end of which, lasting at least $10^{5}$ yrs
\citep{mey03,eld06}, the object is expected to explode as a core-collapse
supernova (CcSN).

Of late, there has been a growing body of observational evidence
that points to LBVs as the {\it direct} progenitors of a subset of
CcSNe. Recently \citet{k06} suggested that LBVs themselves 
might explode; they proposed that the quasi-periodic
modulations observed in the radio light curves of some SNe were 
a manifestation of variable mass loss during the S Doradus phase of
LBVs, during which the star exhibits brightness variations of 
$\sim$1-2 magnitudes. This variability is commonplace amongst 
LBVs \citep{HD94} and is distinct from the giant outbursts such as those of 
P Cygni or $\eta$ Carina.

Interestingly, the progenitor of SN~2006jc was observed
to have a giant eruption just two years prior to explosion
\citep{nak06,fol07,pas07}. \citet{fol07} argue that the progenitor star was of
an early type (WNE)  Wolf-Rayet star. Given that WR stars have never previously
been observed to undergo an LBV-like eruption, \citet{pas07} suggest a massive
binary system consisting of an LBV and a WR star as an alternative progenitor
system for SN~2006jc. In this scenario, the pre-SN explosion coincident with
SN~2006jc is attributed to the LBV, while the SN resulted from the WR
companion. A simpler explanation would be to accept that a single,
presumably massive star exploded during or at the very  end of the LBV phase. 

There are other indications that LBVs may explode. \citet{GY07} 
detected a very luminous source in pre-explosion images of SN~2005gl, 
but it remains to be confirmed whether or not this is a single object. 
Other interesting hints come from the similarities in LBV nebula 
morphologies and the circumstellar medium of SN~1987A \citep{smith07a}.
Finally, the luminous type IIn SN~2006gy, may also have undergone an $\eta$\,Car-type 
eruption prior to explosion \citep{smith07b}.

A potential problem with a scenario in which LBVs explode is that current stellar
evolution models do not predict core-collapse during or soon after the LBV phase;
indeed, in most evolutionary models, the star has not even reached its core He-burning
phase. In other words, in current models, the core is not  evolved enough for
core-collapse to occur. \citet{pas07} point out that known LBVs that have undergone
outbursts still retain some hydrogen and helium while SN~2006jc showed no spectral
features due to hydrogen before $\sim$65\,d. However, it is not inconceivable that the
pre-explosion outburst removed most of the outer layers, and/or that residual amounts
of H or He may not necessarily give rise to strong lines.  

Here, we present the intriguing case of SN~2005gj for which we show
spectroscopic evidence to support the suggestion of \citet{k06} that the
variable mass-loss of S~Dor variables betrays them as the endpoints of very
massive stars. 
\begin{figure*}
\epsfig{file=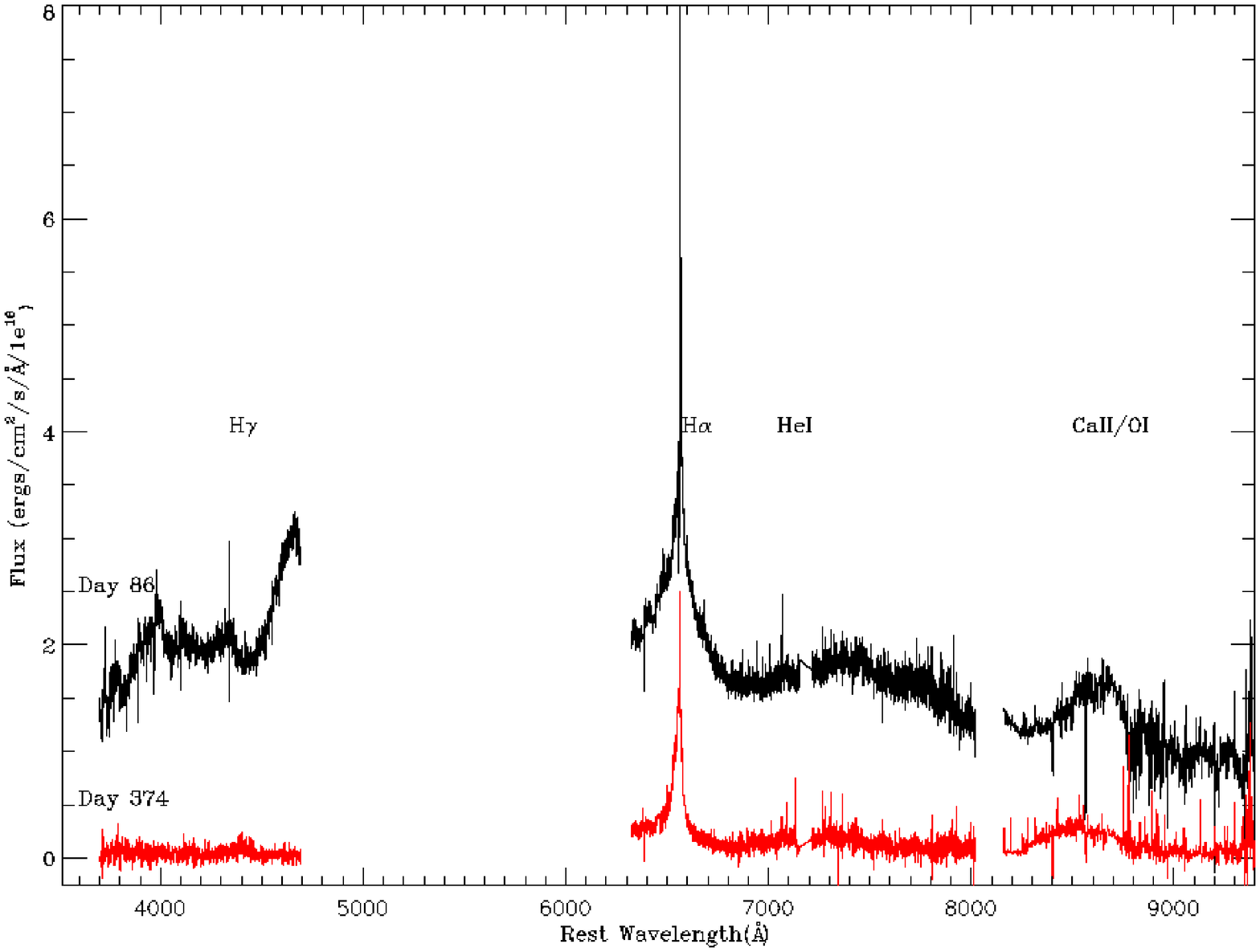, height=120mm, width=80mm, angle=90}
\epsfig{file=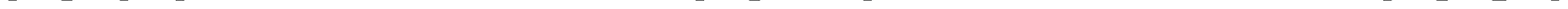, height=50mm,width=38mm,angle=90}
\epsfig{file=ctrundle08_1_fig3.ps, height=50mm,width=38mm, angle=90}
\vspace{-4cm}
\caption[]
{\textbf{Left:} Optical spectra of SN~2005gj at 86 and 374 days after explosion, 
         after correction for the redshift of the host galaxy.
 \textbf{Right:} Multiple component Gaussian profile fits to the H${\alpha}$ profile for both epochs; 
 day 86 in top panel and day 374 in bottom panel (Table~\ref{hafit}). Units of the axes are the
  same as the left panel.}
\label{optspec} 
\end{figure*}

\section{Data acquisition and reduction}
Optical spectra of SN~2005gj were obtained on the 17 Dec. 2005
and 1 Oct. 2006, with the Ultra-Violet Echelle Spectrograph (UVES) 
on the Very Large Telescope. 
These epochs correspond to 86 and 374 days respectively, 
assuming an explosion date of 22 Sep. 2005 \citep{ald06}. 
Dichroic \#2 was used to split the light into the
blue and red arms of the spectrograph. The standard setup with central 
wavelengths at 4370 and 8600 \AA was employed, thus providing coverage 
in the blue of 3760-4985 \AA, and in the red arms of 6705-8520 \AA\ 
and  8660-10255 \AA. For both epochs the target was observed for 2300\,s with a 
slit width of $1\farcs6$ and a seeing of $\sim1\farcs05$  providing 
resolutions of $\sim$6 and 4.5 \kms~in the blue and red arms, respectively.

The data were reduced with the UVES pipeline implemented in the ESO-MIDAS  software.
Wavelength calibration was carried out using Th-Ar arcs taken on  the same night as
the SN~2005gj exposures, while flux calibration and telluric subtraction  of the
data was done with respect to the standard LTT\,1020. The flux calibrated spectra
extracted from each chip and corrected to the rest frame are presented in
Fig.~\ref{optspec}. A redshift of 0.0616 $\pm$ 0.0002, as found by\cite{ald06},  was
adopted as it was consistent with our data. 
\vspace{-0.3cm}
\section{Spectral Analysis}
\label{analysis}
\input{ctrundle08_1_table1.tex}
\vspace{-0.3cm}
The optical spectra of SN~2005gj are dominated by H${\alpha}$ in emission. 
The profile consists of a strong narrow emission component  
superimposed on broad emission wings (Fig.~\ref{optspec}). 
Such H${\alpha}$ profiles are the hallmarks of SN~IIn spectra (`n' standing
for `narrow'), believed to arise from massive progenitors that
have undergone substantial mass-loss prior to explosion \citep{sch90}. 
The multi-component profiles are characterised by broad underlying 
emission due to SN ejecta, an intermediate emission resulting from 
shocked material at the interaction front between the ejecta and the CSM, 
while the narrow component presumably arises from unshocked CSM photoionised by the SN. 

\begin{figure}
\begin{center}
\epsfig{file=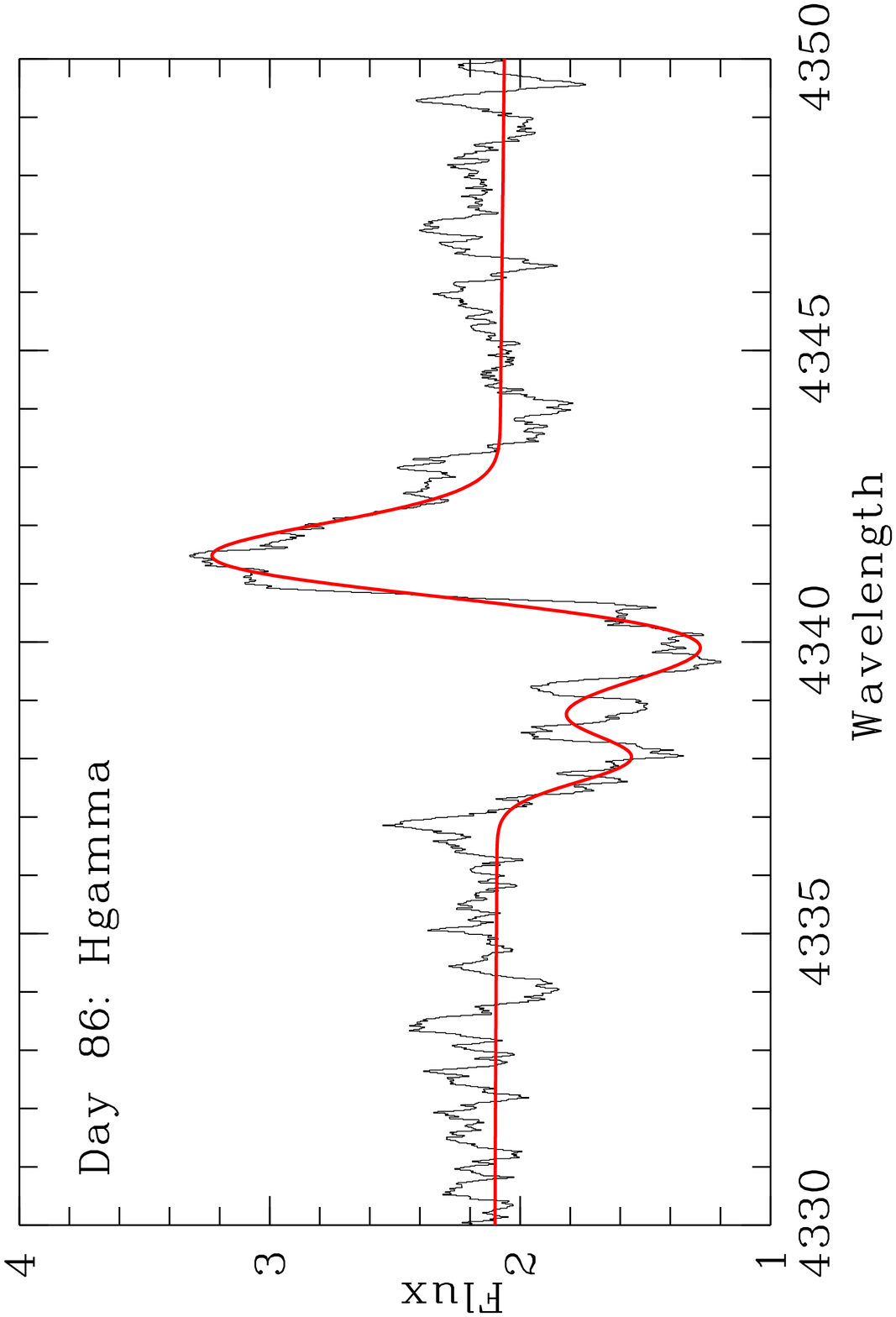, height=43.5mm,width=37mm, angle=270}
\epsfig{file=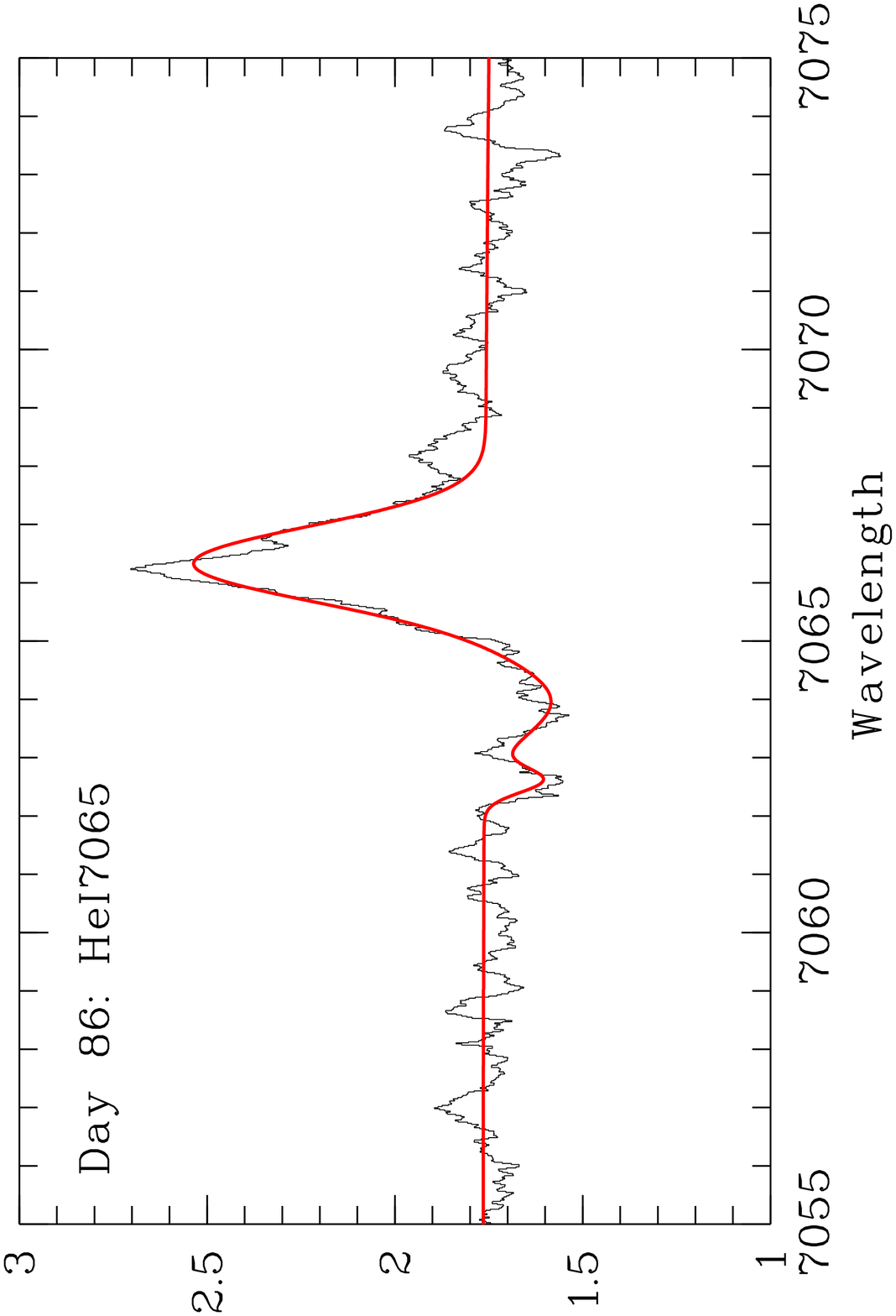, height=43.5mm,width=37mm, angle=270}
\caption[]
{Multiple Gaussian profile fits to the 86\,d spectrum for H$\gamma$ (left),
 and He{\sc i} 7065\,\AA\  (right).The flux on the y-axes are given in units of 
 10$^{-16}$ ergs$^{-1}$ cm$^{-2}$ s$^{-1}$.}
\label{hghe} 
\end{center} 
\end{figure}

A striking feature of the P Cygni-like component of the H$\alpha$ profile at  86\,d
is the presence of two clear absorption troughs (Fig. \ref{optspec}).  This has
previously not been observed in SN spectra.  That this feature is associated with
the SN is clear, given the evolution of the  profile between the two epochs.
\citet{ald06} noted that the absorption component in the narrow P-Cygni profile 
was asymmetric and extended far out towards the blue edge. Due to the higher 
resolution of our spectra \citep[c.f. resolutions of a few hundred
\kms][]{ald06,p07},  we were able to resolve this narrow component.  Careful
inspection of the data revealed clear double absorption troughs in our day 86
spectrum associated with H${\gamma}$, He {\sc i} 7065\,\AA\ (Fig.~\ref{hghe}),  and
also H${\delta}$, and He {\sc i} 6078\,\AA\, albeit with poorer signal-to-noise 
ratios.  By $\sim$400\,d, the H$\alpha$ profile had changed significantly: the
intensity dropped by 17$\times$10$^{-16}$ ergs$^{-1}$ cm$^{-2}$ s$^{-1}$, a single
absorption trough remained, and a broad blue shoulder had developed, probably
corresponding to the emergence of [O {\sc i}].
To obtain quantitative information from these profiles we have decomposed them
with multiple Gaussian profiles for the individual emission and absorption
peaks. The fits are shown in  Figs.~\ref{optspec} \& ~\ref{hghe} and their
physical parameters are summarised in Table~\ref{hafit}. The H${\alpha}$
emission profiles at both epochs were fit by 3 Gaussians representing the
aforementioned broad, intermediate and narrow emission components, however this
intermediate component was not detected in the H${\gamma}$ \&  He {\sc i} 7065
\AA\ profiles. In addition to the  emission peaks, the H$\alpha$ profile from
day 86 was fit with double absorption components, whilst only a single
absorption trough was required for the day 374 spectra. 

Due to our incomplete wavelength coverage, particularly to the blue of
H${\alpha}$, the velocities determined from the Gaussian fits to the broad
emission peak are subject to large uncertainties. From the Balmer lines we
determine an ejecta velocity of 30000 $\pm$ 5000 \kms~at day 86, which is
consistent with that derived by \citeauthor{ald06} from their day 11 spectra.
At day 374 the velocity decreased to 15500 $\pm$ 5000 \kms. The intermediate
H${\alpha}$ emission component was fit by a Gaussian with FWHM 2560 $\pm$ 30
\kms~on day 86 and 1680 $\pm$ 30 \kms~on day 374. Whilst this measurement is a
good indication of the shock velocity for the symmetric profile presented at
day 86, this is not the case for the asymmetric profile on day 374. In this
later spectra the narrow H${\alpha}$ feature is redshifted with-respect-to the
intermediate component, this has been seen in a number of Type~IIn spectra such as
SN~1997ab, 1997eg \& the hybrid Type~Ia/IIn SN~2002ic \citep{k04}. This is
generally attributed to self-absorption in the intermediate component which
preferentially attenuates the red wing of the profile \citep{sal98}. A better
measurement of the shock velocity is therefore the velocity of the blue edge of
the profile at zero intensity, which is 2850 $\pm$ 200 \kms. The luminosities
inferred from the narrow line intensities at day 86 agree well with those
derived by \citeauthor{ald06}, with a significant decline in the luminosities
derived from the narrow component from 1.6$\times$10$^{40}$ ergs$^{-1}$ to
1.8$\times$10$^{39}$ ergs$^{-1}$ by day 374.

Other, less dramatic features in the spectra comprise
a broad feature at $\sim$ 8500 \AA, probably a blend of the Ca
{\sc ii} infrared triplet with O {\sc i} 8446. There is a weak, broad
feature centered on 7300\AA\ which, depending on the SN type, can be
identified as either [Ca {\sc ii}] or [Fe {\sc ii}], with the latter being
typically present in the late-time spectra of SNe Ia.
The [Ca {\sc ii}] lines however, are typically observed in the spectra of Type IIn 
and Ib/c SNe at $\gtrsim$100\,d. 

\section{SN~2005gj: Properties of the unshocked CSM} 
\label{windpar}
These narrow P-Cygni features which are commonly found amongst Type IIn, are
thought to represent the outflow of unshocked CSM surrounding the SN event. As
such, it provides insight into the environment prior to the SN explosion
and on the wind properties of the SN progenitor itself. Stellar wind terminal
velocities, v$_{\infty}$, are usually measured from the blue edge of
strong resonance lines (v$_{\rm edge}$). Unfortunately, there are no
such resonance lines in the optical, and so we will use the Balmer and He {\sc
i} lines to estimate this velocity from their absorption components. As these
lines are not saturated, v$_{\rm edge}$ is likely to underestimate the terminal
velocity \citep{p90}. From the H${\alpha}$ narrow absorption feature, we find
a v$_{\rm edge}$ of 295 \kms~at 86d with an additional component at 120
\kms~whereas the component at 374d has an intermediate velocity of 190 \kms.
The H${\gamma}$ feature at day 86 gives velocities consistent with those found
from H${\alpha}$, however the bluer component of the He {\sc i} profile
appears at a lower velocity with a maximum v$_{\rm edge}$ of 160 \kms. 

We can attempt to derive an estimate of the mass-loss rate from the  H${\alpha}$
feature.  Following \citet{sal98},  ${\rm L}^{\rm Int}_{{\rm H}_\alpha} =
\epsilon_{\rm{H}_\alpha} \dot{M}{\rm v_s^3}/{\rm 4v_w}$ where, $\epsilon_{{\rm
H}_\alpha}$  is an efficiency factor that peaks at $\sim$0.1. L$^{\rm Int}_{{\rm
H}_{\alpha}}$ is the luminosity of the intermediate component, v$_{\rm w}$ is the wind
velocity and v$_{\rm s}$ is the shocked CSM velocity. At day 86, this implies \.M =
6.4 and 2.6 $\times10^{-2}$  M$_\odot$ yr$^{-1}$ from the high (295 \kms) and low
(120 \kms) velocity components. At day 374, the derived \.M is 1.7$\times10^{-2}$
M$_\odot$ yr$^{-1}$ using  v$_{\rm w}$ = 190 \kms and  v$_{\rm s}$ = 2850 \kms, one
should note that the shock velocity is a little less certain in this case (see
Sect.~\ref{analysis}). These are very large mass-loss rates but are typical of those
derived from the H$\alpha$ luminosity in supernovae which show strong interaction
between the ejecta and CSM \citep[e.g. SNe~1997ab, 2002ic][]{sal98,k04}.  However,
\citeauthor{k04} found that mass-loss rates derived from dust  measurements in the
infrared are a magnitude lower. A similar observation can be made by comparing
the mass-loss rates of SN~2005gj derived here and that of \citeauthor{p07} from x-ray
luminosities. Several caveats must be borne in mind when estimating mass loss rates
via  H$\alpha$ profiles. Two of the most serious shortfalls involve the unknown
ionisation conditions and uncertainties in $\epsilon_{\rm{H}_\alpha}$. Realistic
constraints can only be derived from detailed modelling  which is beyond the scope of
this work.
\begin{figure}
\begin{center}
\epsfig{file=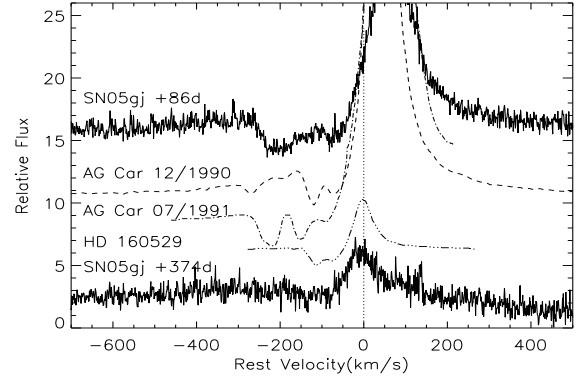, height=85mm,width=55mm, angle=90}
\caption[]
{Comparison of the H$\alpha$ profile of SN~2005gj in velocity space with that of the
luminous blue variables, AG Car \& HD 160529. Data taken from \cite{stahl01,stahl03}.}
\label{havspec}  
\end{center} 
\end{figure}
\section{Discussion}
\label{discussion} 

SN~2005gj has been compared to the prototype for the hybrid Type~Ia/IIn objects,
SN~2002ic \citep{ald06}. SN~2002ic was initially classified as a Type Ia, due to the presence of S {\sc
ii} lines and a blue shifted absorption feature of Si {\sc ii} 6335 \AA, but was quickly
reclassified as a hybrid Type~Ia/IIn due to the presence of multi-component Balmer lines
signifying interaction with the surrounding circumstellar medium \citep{h03}.  This was an
exciting discovery as the presence  of circumstellar hydrogen in the spectra of a type Ia
supernovae would shed light on the nature of Type Ia progenitors.  One scenario put
forward by \citeauthor{h03} is that the dense CSM could represent mass-loss from
an asymptotic-giant-branch (AGB) star. Unfortunately, the true classification and
interpretation of SN~2002ic remains under debate. In particular, \citet{b06} argue that it
can equally  well be compared to a core-collapse SN, such as the type Ic SN~2004aw. This
comparison is most convincing at late epochs, in fact the resemblance of SN~2002ic to type IIn
SNe had been noted earlier \citep[e.g.][]{h03}.

In spite of the H$\alpha$ feature indicating CSM interaction, the early time optical
spectra of SN~2002ic were undeniably Ia-like \citep[see Fig.~3][]{h03}.  
\citet{ald06} \& \citet{p07} argue that SN~2005gj could be compared to the 
overluminous Type Ia, SN~1991T `diluted' by the presence of significant CSM.  However,
the typical Ia spectral features due to S {\sc ii} and Si {\sc ii} which  formed the
crux of the Ia-evidence for SN~2002ic, are barely discernible in the  spectra of
SN~2005gj. They attribute this to stronger interaction of the SN~2005gj with the CSM
than for SN~2002ic, arguing that this claim is supported by its broader light curve
and higher H$\alpha$ luminosities.

On the basis of the data presented here, we cannot verify or exclude the type Ia
identification of SN~2005gj. However, we present an alternative scenario that is
consistent with the observations. We focus our attention on the double absorption troughs
seen in the day 86 spectrum.  Only in one other SN spectra, SN~1998S, have double
absorption troughs been dectected \citep{bowen00,fassia01}. However the shape of the lines
differ significantly. In SN~1998S a narrow P-Cygni profile of $\sim$50 \kms is
superimposed on a broad, shallow absorption profile \citep[see Fig.~8]{bowen00}.
Additionally, this absorption decreases in relative intensity from lower to higher Balmer
lines, suggestive of an optical depth dependence that is inconsistent with the SN~2005gj
spectra. \citet{chugai02} invoke several mechanisms that could accelerate the
circumstellar gas around SN~1998S. These models have been tailored to match the early-time
spectra of SN~1998S and it is difficult to see how these might be tailored to the
significantly later epochs of our data. This type of test may provide some insight but is
beyond the scope of this letter.

Here we would like to stress the likeness of the profiles in SN~2005gj with another
stellar object: Luminous Blue Variables. Multiple-absorption component
H${\alpha}$ profiles have been detected in  many LBVs, {\it viz\/} AG Carinae, R127, R81,
HD160529, R66, \& P-Cygni \citep{wolf81,stahl83,stahl83a,stahl01,stahl03,l94}. Fig. 3
shows several examples of the H$\alpha$ profiles of LBVs gleaned from the literature.  In
LBVs these multiple-absorption component profiles signify their history of episodic
mass-loss events. \cite{v02} interpreted the variable mass-loss and wind velocity
behaviour of LBVs as being due to a change in the ionization of the dominant wind driving
ion Fe (wind bistability). This mechanism was also invoked to explain the variations in
the radio lightcurves of  transitional SNe (Kotak \& Vink 2006). However, regardless of
the mechanism the resemblance of the double-troughed LBV H$\alpha$ profiles to that of
SN~2005gj remains.

An interesting hydrodynamical simulation by \citet{vm07}
shows the evolution of CSM around a 60M$_\odot$ star as it evolves from
main-sequence to core-collapse via an LBV and WR phase. This simulation shows the
formation of multiple absorption components from CSM shells formed by a fast WR wind
running into an LBV shell. However these multiple shells occur in the very early stage
of the WR-lifetime and hence if such a scenario was implied for SN~2005gj, its WR-phase
would have been short-lived.

While the morphological resemblance of the line profiles observed in SN~2005gj and
in LBVs is impressive enough, an even more compelling similarity is that of LBV
wind velocities with those derived from the absorption components of SN~2005gj. LBV
velocities lie in the $\sim$50--300\,\kms range, in strikingly good agreement with
those derived from SN~2005gj (Table 1), which are larger than those observed
in AGB winds ($\sim$10\,\kms) and red-supergiant winds and much lower than typical
WR-star winds. Typical timescales of S Doradus mass-loss events in LBVs are on the
scale of years up to $\sim$100 yrs \citep{v01}. From a simple calculation we can
determine the timescales of the profile variations detected in SN~2005gj.
Using an intermediate ejecta velocity of 22500 \kms, a time variation of 288 days
and an edge velocity of 300 \kms~we get P$_{\rm w}$=R$_{\rm ejecta}$/v$_{\rm edge}$
$\sim 60$ yrs. Thus the timescales associated with mass-loss events in LBV and SN
progenitors are consistent. Note the 60~yr timescale is only an upper limit as we
have not obtained a well-sampled temporal dataset of high-resolution spectra. Hence
the variations are likely to be on shorter timescales still consistent with S
Doradus variations.  Even the change into a single absorption component (374\,d)
mimics the behaviour of LBVs such as AG~Car which show similar variations on
timescales of years (Fig.~3).

For SN~2005gj the line profiles, wind velocities and wind periodicity are remarkably
reminiscent of LBVs and provide significant support for a direct link between supernovae
and luminous blue variables.  This may have a huge impact on our understanding of massive
star evolution since current theory predicts that LBVs, which are in a short-term,
intermediary phase of evolution, should not undergo core-collapse.
\vspace{-0.3cm}
\section{Acknowledgements}  
The authors acknowledge fruitful discussions with L. Dessart
\& N. Langer.
\vspace{-0.4cm}
\bibliography{ctrundle08_1}  
\end{document}

%% file: ctrundle08_1_table1.tex
\begin{center}
\begin{table}
\caption[Decomposition fits to H$_\alpha$]
{\label{hafit} Parameters of the gaussian fits to the narrow components of H$_{\gamma,\alpha}$  \& He {\sc
i} 7065 \AA\  lines of SN 2005gj at 86 and 374 days after explosion. `Narrow E' - the narrow emission component,
`NAC (a)' and `NAC (b)' - the blue and red absorption components respectively. Each profile has been fit by multiple gaussian
profiles, representing the broad, intermediate and narrow components of the profile.
 }
\vspace{-0.5cm}
\begin{flushleft}
\centering
\begin{tabular}{lccc} \hline \hline
\\
Fit properties              & Narrow E & NAC (a) &  NAC (b) \\           
\hline
                  & \multicolumn{3}{c}{Day 86: H$_\alpha$}           \\
\hline
\\
$\lambda$ (\AA)             &6564.29$\pm$0.02  & 6558.35$\pm$0.06& 6561.25$\pm$0.06\\  
FWHM (\kms)                 &  128  $\pm$ 2    &     70$\pm$6	 &  133$\pm$21     \\  
L (10$^{40}$ ergs s$^{-1}$)&1.61$\pm$0.02     &   0.14$\pm$0.01&  0.20 $\pm$0.02  \\  
v$_{\rm edge}$ (\kms)       &        & 295 $\pm$ 57	 &  120 $\pm$ 75   \\  
\hline

                  & \multicolumn{3}{c}{Day 86: H$_\gamma$}           \\
\hline
$\lambda$ (\AA)             &4341.39$\pm$0.02  &4338.01$\pm$0.10 & 4340.04$\pm$0.10  \\ 
FWHM (\kms)                 &  97  $\pm$ 3     &    68  $\pm$ 5  &   115  $\pm$ 4    \\ 
L (10$^{40}$ ergs s$^{-1}$)& 0.15$\pm$0.04    &  0.04$\pm$0.01 &  0.12$\pm$0.05   \\ 
v$_{\rm edge}$  (\kms)     &         &   286 $\pm$ 57  &  115  $\pm$ 64    \\ %
\hline			    		      
                  & \multicolumn{3}{c}{Day 86: He{\sc i}7065}           \\
\hline
$\lambda$ (\AA)             &7066.33$\pm$0.03 & 7062.61$\pm$0.09& 7063.97$\pm$0.10  \\ 
FWHM (\kms)                 &  64 $\pm$ 3     &     22 $\pm$9	&     62$\pm$15     \\ 
L (10$^{40}$ ergs s$^{-1}$)&0.10$\pm$0.02    &  0.006$\pm$0.001&   0.022$\pm$0.004 \\ 
v$_{\rm edge}$  (\kms)      &        &   160$\pm$ 59	& 95  $\pm$ 57      \\ 
\hline			       	       
                  & \multicolumn{3}{c}{Day 374: H$_\alpha$}          \\

\hline
$\lambda$ (\AA)              & 6562.60$\pm$0.04&6559.70 $\pm$0.12&		     \\ 
FWHM (\kms)                  &  88$\pm$ 10     &  110 $\pm$ 23   &		     \\ 
L (10$^{40}$ ergs s$^{-1}$) & 0.18$\pm$ 0.09  & 0.028$\pm$ 0.005&		     \\ 
v$_{\rm edge}$  (\kms)       &        & 190 $\pm$ 64	 &		     \\ 
\hline			       	       
\end{tabular}
\end{flushleft}
\end{table}
\end{center}